\documentclass[a4paper,fleqn,usenatbib]{mnras}

\usepackage[T1]{fontenc}
\usepackage{ae,aecompl}
\usepackage{hyperref}

\usepackage{graphicx}    % Including figure files
\usepackage{amsmath}    % Advanced maths commands
\usepackage{amssymb}    % Extra maths symbols

\title[G279.0+1.1: a new extended source of gamma rays]{G279.0+1.1: a new extended source of high-energy gamma rays}

\author[Araya]{
M. Araya$^{1}$
\\
$^{1}$Centro de Investigaciones Espaciales and Escuela de F\'isica, Universidad de Costa Rica}

\date{Accepted 2020 January 22. Received 2020 January 22; in original form 2019 June 21}

\pubyear{2019}

\begin{document}
\label{firstpage}
\pagerange{\pageref{firstpage}--\pageref{lastpage}}
\maketitle

\begin{abstract}
G279.0+1.1 is a supernova remnant (SNR) with poorly known parameters, first detected as a dim radio source and classified as an evolved system. An analysis of data from the \textit{Fermi}-LAT revealing for the first time an extended source of gamma rays in the region is presented. The diameter of the GeV region found is $\sim 2.8\degr$, larger than the latest estimate of the SNR size from radio data. The gamma-ray emission covers most of the known shell and extends further to the north and east of the bulk of the radio emission. The photon spectrum in the 0.5--500 GeV range can be described by a simple power law, $\frac{dN}{dE} \propto E^{-\Gamma}$, with a spectral index of $\Gamma = 1.86\pm 0.03_{stat} \pm 0.06_{sys}$. In the leptonic scenario, a steep particle spectrum is required and a distance lower than the previously estimated value of 3 kpc is favored. The possibility that the high-energy emission results from electrons that already escaped the SNR is also investigated. A hadronic scenario for the gamma rays yields a particle spectral index of $\sim2.0$ and no significant constraints on the distance. The production of gamma rays in old SNRs is discussed. More observations of this source are encouraged to probe the true extent of the shell and its age.
\end{abstract}


\section{Introduction} \label{sec:intro}
A supernova remnant (SNR) is created by an explosion which could be caused by the death of a massive star or by the destruction of a white dwarf (or a binary white dwarf system). The shocks of these explosions are capable of accelerating charged particles to very high energies, as shown historically by theoretical work and observations from radio to X-rays \citep[e.g.,][]{1978MNRAS.182..147B,1978ApJ...221L..29B,1981ApJ...245..912R,1995Natur.378..255K}. More recently, high-energy (GeV) and very-high energy (TeV) data from different observatories have established SNRs as gamma-ray sources \citep[e.g.,][]{2016ApJS..224....8A,2018A&A...612A...1H,2013Sci...339..807A,2016ApJ...816..100J}, though a definitive prove that SNRs can accelerate particles to the highest energies seen in Galactic cosmic rays is still lacking \citep[see, e.g.,][]{2013A&ARv..21...70B}.

The gamma ray spectra of SNRs are diverse and likely highly depend on the environment in which the expansion takes place \citep[e.g.,][]{2012ApJ...761..133Y,Yasuda_2019,2019MNRAS.487.3199C}. Some middle-aged SNRs interact with molecular clouds or high-density gas and show bright gamma-ray emission that is consistent with a hadronic scenario. Young SNRs such as Cas A with intermidiate gas densities ($\sim 1$ cm$^{-3}$) show spectra consistent with a mixture of hadronic and leptonic emission \citep[e.g.,][]{2010ApJ...720...20A}. Other, relatively young SNRs such as RX J1713.7-3946, seem to be located in low density environments, show considerable non-thermal X-ray fluxes, sometimes relatively low radio emission, and have a hard spectral energy distribution (SED) at GeV energies. Such a spectrum is expected from inverse Compton (IC) emission from high-energy electrons \citep{2012ApJ...761..133Y}, although for RX J1713.7-3946 there are problems with the single zone leptonic scenario and hadronic models are also able to account for the gamma rays \citep{2019MNRAS.487.3199C}.

G279.0+1.1 was first identified in radio observations by \cite{1988MNRAS.234..971W} as the shell of an SNR with an angular diameter of $1.6^{\circ}$. They estimated a distance to the source of 3 kpc and pointed out that the SNR is probably old based on its low surface brightness. They also noted there are small, faint CO clouds around the SNR with a velocity close to 0 km s$^{-1}$ which would place them close to the Carina arm, and thus could be associated to the SNR. According to \cite{1995MNRAS.277..319D}, there are three old pulsars within $2.6\degr$ of the SNR centre: PSR 0953-52, PSR 0959-54 and PSR 1014-53, and they point out that PSR 0953-52 is the closest one to the SNR as seen in the sky. It has a dispersion measure distance that the authors estimated to be 5 kpc, but with large uncertainty. However, the ATNF Pulsar Catalogue \citep{2005AJ....129.1993M}\footnote{See also http://www.atnf.csiro.au/research/pulsar/psrcat/} gives revised distances to these pulsars of 0.4 kpc, 0.3 kpc and 0.12 kpc, respectively. These pulsars have estimated ages with orders of magnitude of $10^5 - 10^6$ yr. The pulsar that is seen closest to the SNR centre, at $0.64\degr$, is PSR 0953-52. Given its old age of 3.9 Myr \citep{1993ApJS...88..529T}, \cite{1995MNRAS.277..319D} argue that PSR 0959-54, located 1.6$^{\circ}$ from the SNR centre and with an estimated age of 0.5 Myr \citep{1993ApJS...88..529T}, is more likely than PSR 0953-52 to be associated to the SNR. Seven additional pulsars are found in the ATNF catalogue within $1.15\degr$ and $2.65\degr$ of the centre of the SNR \citep{2005AJ....129.1993M}. These pulsars are all quite evolved (with ages in the range $10^5 - 10^6$ yr), except for PSR J0940-5428, seen $2.9\degr$ away from the SNR centre, with an estimated age of 40 kyr. Of all the ten pulsars known within $3\degr$ of the SNR, this is also the pulsar with the highest spin-down power ($\dot E = 1.9\times 10^{36}$ erg s$^{-1}$). It is associated to the gamma-ray source 4FGL J0941.1-5429 \citep{2019arXiv190210045T} and it is seen to the south west of the SNR well outside its radio shell.

\cite{1995MNRAS.277..319D} presented higher resolution radio observations of G279.0+1.1 showing a well-defined ring of emission consisting also of possible radio filaments (later confirmed in images by \cite{1996A&AS..118..329W}). They measured a flux density index of $\alpha = 0.6\pm 0.3$ ($S_{\nu} \propto \nu^{-\alpha}$) and large amounts of polarized emission (up to 50\%) with a significant degree of uniformity in both the Faraday rotation measures and the magnetic field structures around the SNR. They argued that this is typical of old remnants having less turbulence as their shock slows down. Similar properties are seen for G116.5+1.1 \citep{1981A&A....99...17R}. According to \cite{1995MNRAS.277..319D}, the variation of percentage polarization across G279.0+1.1 may be caused by interactions with an inhomogeneous interstellar medium. They also showed an image with the locations of the closest CO clouds discussed previously, with kinematic distances consistent with 3 kpc. These are a small eastern cloud and a western cloud (see their Fig. 8). However, no OH maser line (1720 MHz) emission, characteristic of shock and molecular cloud interactions, was detected from G279.0+1.1 in a survey of maser emission \citep{1997AJ....114.2058G}.

More recently, \cite{2009MNRAS.394.1791S} detected H$\alpha$ filaments in part tracing the radio emission of G279.0+1.1. The optical filaments have spectral features consistent with an origin from an SNR such as the intensity ratio of the SII and H$\alpha$ lines. The authors also derived a large size for the source from their estimated distance of 3 kpc, which is consistent with an evolved SNR. However, they noted that this value is higly uncertain. Based on their observations and an inspection of data from the SUMSS radio survey at 843 MHz \citep{2003MNRAS.342.1117M} they concluded that the angular diameter of the remnant is in fact $2.3^{\circ}$, larger than that originally determined.

No dedicated X-ray studies have been carried until now on G279.0+1.1. No sources are found in the region of the SNR at hard X-rays in the 70 month \textit{Swift}-BAT survey \citep{2013ApJS..207...19B}

The \emph{Fermi} Large Area Telescope (LAT), a converter/tracker telescope with sensitivity in the energy range between 20 MeV and $> 500$ MeV \citep{2009ApJ...697.1071A}, has expanded our knowledge of SNR emission and it has cataloged several dozen SNRs \citep{2016ApJS..224....8A}. In the latest LAT catalog, the LAT 8-year Point Source Catalog \citep[4FGL,][]{2019arXiv190210045T}, a point source candidate (i.e., with a detection significance below the 5$\sigma$ level) is found in the southwest of G279.0+1.1, labeled 4FGL J0951.7-5328. The 4FGL catalog indicates that this source candidate might be associated to the SNR. It has a power-law gamma-ray spectrum with an index of $2.12\pm 0.19$. No other gamma-ray sources in the region of the SNR are found in recent catalogs such as the The Third Catalog of Hard Fermi-LAT Sources \citep{2017ApJS..232...18A} nor the H.E.S.S. Galactic plane survey \citep{2018A&A...612A...1H}.

In this work, a detailed analysis of gamma-ray data from the \textit{Fermi}-LAT is carried out in the region of G279.0+1.1 and the discovery of extended GeV emission at the location of the SNR is reported. The gamma-ray source has a size that is larger than the previously reported sizes obtained from optical and radio data but it might be consistent with recent radio observations. The implications of the gamma-ray emission properties for this SNR are examined.

\section{\emph{Fermi} LAT data} \label{sec:LAT}
We analized Pass 8 data from the beginning of the mission (August 2008) to April 2019 with the most recent response functions (P8R3) and the standard publicly available software \textit{fermitools}, version 1.0.1, applying recommended cuts. Events falling in the SOURCE class (filtered with the parameter evclass=128) were used with event types including front and back interactions (using evtype=3), and in the reconstructed energy range 0.5--500 GeV. They were selected having a maximum zenith angle of 90$\degr$, to avoid contamination from Earth's limb, for time intervals when the data quality was good. The total livetime of the observation amounts to $\sim 8.9$ years after cuts. A spatial binning scale of 0.1$\degr$ per pixel was used and the exposure was calculated using ten logarithmically spaced bins per decade in energy. The region of interest analyzed had a radius of 20$\degr$ around the coordinates (J2000) RA=$150\degr$, Dec=$-53\degr$. The maximum likelihood technique \citep{1996ApJ...461..396M} was used to maximize the probability for the model to explain the data, fitting the free parameters of the sources in the model. These sources were taken from the latest 4FGL LAT catalog, as well as the Galactic diffuse emission, given by the file gll\_iem\_v07.fits, and the residual background and extragalactic emission, given by the file iso\_P8R3\_SOURCE\_V2\_v1.txt. The detection significance was estimated with the test statistic (TS), defined as $-2\times$log$(L_0/L)$, with $L$ and $L_0$ the values of the maximum likelihoods for nested models (or for example, one having an additional source and the other being the null hypothesis). The source candidate 4FGL J0951.7-5328 was not included in the model in order to carefully study the emission in the region. The source candidate 4FGL J0957.6-5208 as well as the dim source 4FGL J1004.6-5215 are located in the region and have no association to known objects. Both were also removed from the model.

\begin{figure}
\includegraphics[width=10cm,height=8cm]{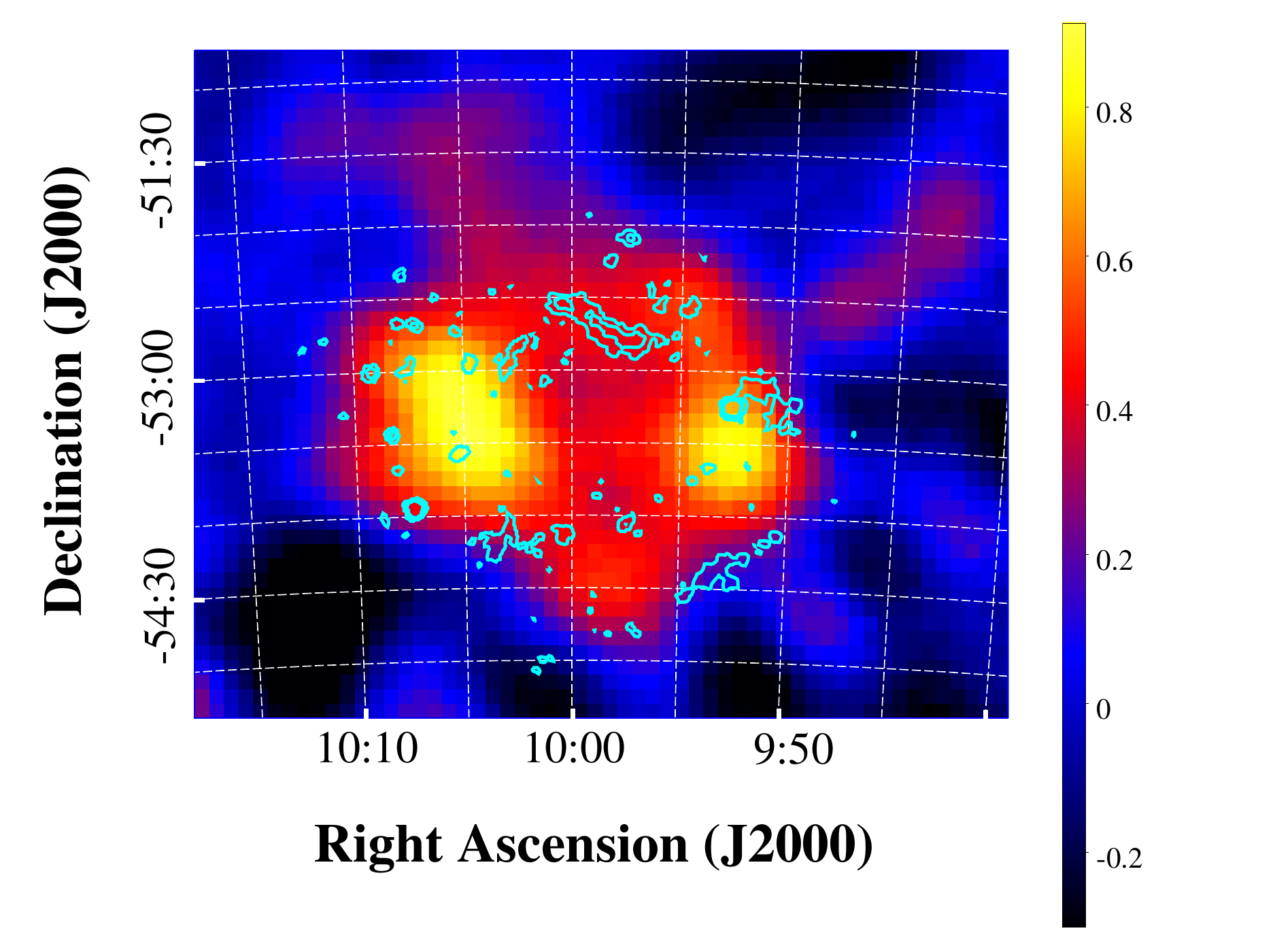}
\caption{Residuals count map of gamma-ray emission above 5 GeV obtained after subtracting the background model from the data and smoothing with a Gaussian function with $\sigma=0.3\degr$. The contours show the radio emission from a 4850 MHz continuum observation \citep{1993AJ....106.1095C} at the 0.025, 0.0625 and 0.1 janskies/beam levels. \label{fig1}}
\end{figure}

An initial inspection of the data is done by fitting events in the energy range 5-500 GeV, leaving the normalization of the sources free in a radius of 14$\degr$ around the centre to improve the quality of the residuals, which were obtained after subtracting the best-fit model to the data. The fit was improved by first adding three new point sources to the model where residuals were seen. Whenever a point source was added to the model, the tool \emph{gtfindsrc} was used to optimize its location. The added sources are located far away ($> 8\degr$) from the centre of the region and have TS values of $\sim 20-70$. Another interesting region of excess emission was seen around the coordinates (J2000) RA=$152.6\degr$, Dec=$-58.1\degr$, close to the very large unidentified source FGES J1036.3-5833 and Westerlund 2 \citep{2017ApJ...843..139A}. Different models were investigated for this excess emission such as a combination of point sources and a  disc. In the end a disc with a radius of $1.3\degr$ was chosen as the best representation. The resulting TS of this template was 62.9 above 5 GeV with a spectral index of $1.8\pm0.1_{stat}$. Since this might be an interesting new extended source with a hard spectrum, a detailed analysis is left for future work. This emission is located $\sim5\degr$ from the centre of the region of interest and the choice of morphology used was seen to have no significant effects in the results of this work. Finally, with the improved model new residuals were obtained as shown in Fig. \ref{fig1}. A clear emission excess is seen at the location of G279.0+1.1 and the results of the morphological analsis are shown next.

\subsection{Morphology of the gamma-ray emission}
Taking advantage of the improved resolution at higher energies the morphology of the source above 5 GeV was studied by comparing the fits to different spatial templates: combinations of point sources, a disc, a gaussian and a ring.

The sizes and locations of the disc, ring and gaussian templates were changed in a grid. The locations of the templates were varied in a $3\times3\degr$ region around the radio centre of G279.0+1.1 in steps of $0.1\degr$, while their extensions were changed from $0.5\degr$ to $2\degr$ in steps of $0.1\degr$ at each position. At each step, a maximum likelihood fit is performed starting with the best-fit parameters of the background sources described earlier, keeping the normalization free for sources located at distances of less than $5\degr$ from the centre of the region. On the other hand, a TS map was calculated to study the point source hypothesis. The TS values were calculated by fitting the parameters of a point source in a grid of positions within a $4\times4\degr$ region around the centre of the SNR. Three locations are found with the highest TS values of 11.2, 11.8 and 17.1. These values are not significant enough to claim the existence of new point sources. On the other hand, this situation is typically seen in point-source TS maps of an extended source \citep[e.g.,][]{2017ApJ...843..139A}. A simple power-law spectrum was assumed to fit all the spatial models, and this is shown later on to be the preferred spectral shape. Table \ref{tab1} presents a summary of the best-fit parameters obtained in each case. For the different morphological models the value of the Akaike Information Criterion \citep[AIC,][]{1974ITAC...19..716A} is shown. The AIC is defined as $2k -2\, \mbox{log}L$, where $k$ is the number of parameters in the model. The best hypothesis is the one that minimizes the AIC.

\begin{table*}
\centering
\caption{AIC values above 5 GeV for various morphological hypotheses.}
\begin{tabular}{|c|c|c|c|}
\hline
\hline
\textbf{Spatial model} & \textbf{Fitted size$^{a}$ ($\degr$)} & \textbf{AIC} & \textbf{D.o.f.$^{b}$} \\
\hline
Uniform disc & $1.44\pm 0.10 \degr$ & 256948.7 & 5 \\
\hline
2D Gaussian & $1.0\pm 0.2 \degr$ & 256967.4 & 5  \\
\hline
%%%el error de los radios ya esta arreglado
Ring &  $1.44\pm 0.08 \degr$  & 256942.4 & 6 \\
&  $0.3 \pm 0.2 \degr$  &  & \\
\hline
3 point sources & -- & 257086.7 & 12 \\
\hline
\end{tabular}\\
\textsuperscript{$a$}\footnotesize{Radii for the disc and ring and sigma for the gaussian. Uncertainties calculated at the 3$\sigma$-level.}\\
\textsuperscript{$b$}\footnotesize{Additional number of free parameters with respect to the null hypothesis.}\\
\label{tab1}
\end{table*}

With a TS of 181.0, the ring template is slightly preferred with respect to the disc (with a TS of 172.5) at a level close to 3$\sigma$ for one additional degree of freedom. The ring template results in the lowest AIC as well and it was chosen as the best morphological representation of the data among these models for the rest of the analysis. The ring is centered at the coordinates (J2000) RA=$150.0\pm0.2 \degr$, Dec=$-53.2^{+0.2}_{-0.3} \degr$, with inner and outer radii of $1.44 \pm 0.08 \degr$ and $0.3 \pm 0.2 \degr$, respectively (3$\sigma$-level uncertainties). The location and outer radius of the ring are consistent with the location and size found for the best fit disc template.

\subsection{Spectrum in the 0.5-500 GeV range}
The spectrum of the source was studied for the whole energy range (0.5-500 GeV) using the best-fit spatial template found in this section, and the results are shown in Table \ref{tab2}. The gamma-ray spectrum of G279.0+1.1 is best described by a simple power-law. The detection significance of the source in this energy range is very high ($\sim 21 \sigma$). The spectral index and integrated flux (0.5-500 GeV) are $1.86\pm 0.03_{stat} \pm 0.06_{sys}$ and $(7.5\pm0.5_{stat}\pm 1.6_{sys})\times 10^{-9}$ photons cm$^{-2}$ s$^{-1}$, respectively. The model of the Galactic diffuse emission is the main source of systematic uncertainty for faint sources \citep{2019arXiv190210045T}. This uncertainty was estimated by performing a fit with several alternative emission models as done by \cite{2016ApJS..224....8A}. The results were validated with the eight alternative models described by \cite{2016ApJS..224....8A}. The detection significance of the source was always above $20 \sigma$ for these models. The uncertainties in the effective area are propagated onto the spectral index and flux with the use of a set of bracketing response functions as recommended by \cite{2012ApJS..203....4A}. The quoted systematic uncertainty is the combined effect of these two independent factors.

\begin{table*}
\centering
\caption{TS values$^{a}$ for different spectral shapes (0.5-500 GeV).}
\begin{tabular}{|c|c|}
\hline
\hline
\textbf{Spectral shape} & \textbf{TS} \\
\hline
Simple power-law  & 454.5 \\
\hline
LogParabola  & 457.6 \\
\hline
Power-law with & 368.1 \\
exponential cutoff &   \\
\hline
\end{tabular}\\
\textsuperscript{$a$}\footnotesize{Calculated with respect to the null hypothesis.}\\
\label{tab2}
\end{table*}

\subsection{Spectral energy distribution and multiwavelength observations}
We binned the LAT data in nine intervals in the 0.5-500 GeV range to obtain spectral points along with the global fit. In each interval we fixed the spectral index of the source to the value found above and fit the normalization. The source was detected above $5\sigma$ in the first seven bins and above $3\sigma$ in the last two. The flux values are shown in Table \ref{tab3}.

\begin{figure}
\includegraphics[width=10.5cm,height=9cm]{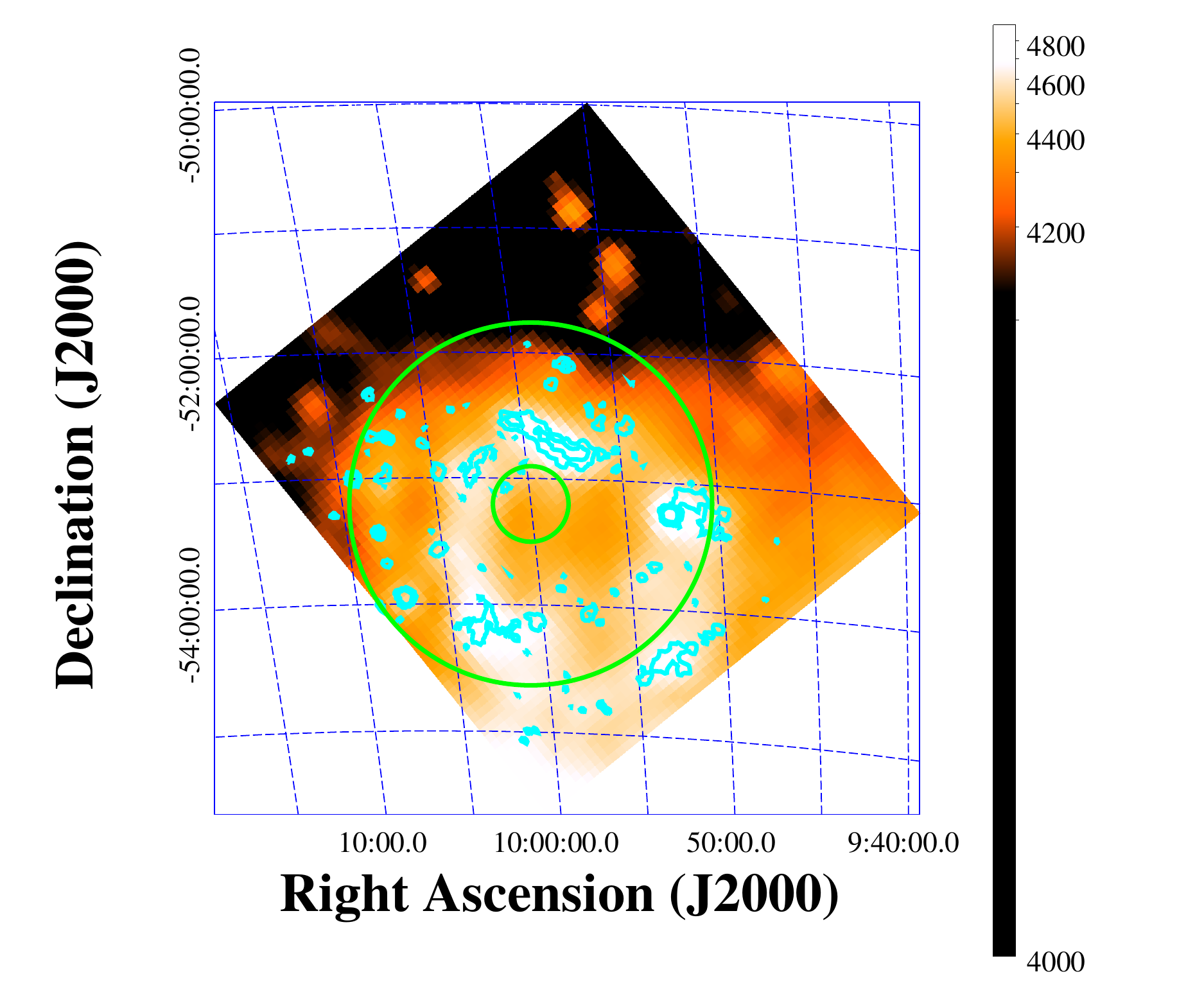}
\caption{CHIPASS 1.4 GHz radio continuum map (in units of mK) with the same 4850 MHz contours shown in Fig. \ref{fig1}. The green circles represent the inner and outer radii of the ring modeling the gamma-ray emission found in this work. The radio ring can be seen in coincidence with the bulk of the 4850 MHz emission. The color scale was saturated to highlight possible structures of the shell to the north and east of the radio ring.\label{fig2}}
\end{figure}

Fig. \ref{fig2} shows a CHIPASS 1.4 GHz radio continuum map \citep{2014PASA...31....7C} of the G279.0+1.1 region where the radio (ring-like) shell of the SNR is seen. The inner and outer bounds of the best-fit ring template obtained from the gamma-ray data are shown for comparison. The gamma-ray region is slightly larger than the radio ring and its centre is displaced to the northeast. This recent radio map also shows emission extending towards the east and north of the ring which might be part of the shell of the SNR. This is supported by the results presented here. There are no H II regions near the SNR found in the WISE Catalog of Galactic H II Regions \citep{2014ApJS..212....1A} that could be responsible for the radio emission to the north. Deeper radio observations as well as detailed future studies of the gas in the region will be necessary to confirm or reject this possibility.

To study the multiwavelength spectrum an upper limit on the TeV emission by H.E.S.S. was used \citep{2018A&A...612A...3H}, and radio fluxes were taken from \cite{1988MNRAS.234..971W} and \cite{1995MNRAS.277..319D}. The radio fluxes reported in the literature come mainly from the brightest parts of the radio shell, which is a region slightly smaller than the gamma-ray emission region found here.

%%%%ESTA TABLA YA CONTIENE LOS VALORES DE LA SED DEL ANILLO
\begin{table*}
\centering
\caption{Gamma-ray fluxes of the SNR G279.0+1.1.}
\begin{tabular}{|c|c|}
\hline
\hline
\textbf{Energy range (GeV)} & $E^2\frac{dN}{dE}$ \textbf{($\times 10^{-12} $ erg cm$^{-2}$ s$^{-1}$)} \\
\hline
0.5--1.0  &  $8.5\pm 0.9$\\
1.0--2.0  &  $4.4\pm 0.9$\\
2.0--4.0  &  $6.4\pm 1.0$\\
4.0--8.0  &  $7.6\pm 1.2$\\
8.0--16.0  &  $7.7\pm 1.4$\\
16.0--32.0  &  $8.6\pm 1.7$\\
32.0--64.0  &  $10.0\pm 2.2$\\
64.0--128.0  &  $7.1\pm 2.6$\\
128.0--500.0 &   $12.7\pm 4.3$\\
\hline
\end{tabular}\\
\label{tab3}
\end{table*}

\section{Multiwavelength models and discussion}
Two scenarios for the origin of the high-energy radiation were considered to constrain some of the poorly known physical parameters of G279.0+1.1. In the first scenario the gamma rays are produced by IC emission from a population of high-energy electrons interacting with the cosmic microwave background (CMB) and in the second one, a one-zone hadronic model was applied.

\begin{figure}
\includegraphics[width=9cm,height=5cm]{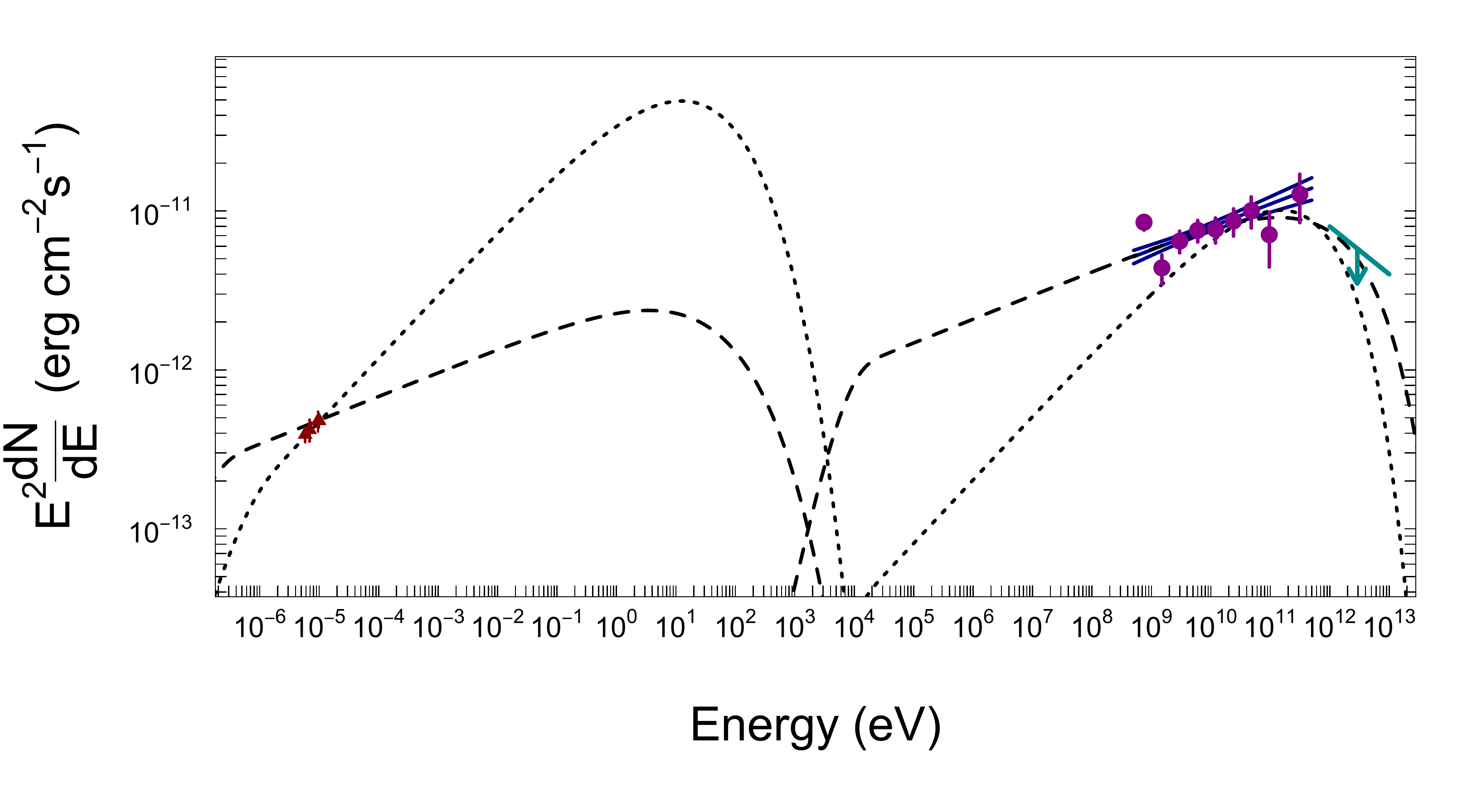}
\caption{Multiwavelength SED of G279.0+1.1 with the leptonic models described in the text adopting different slopes in the radio flux density ($S_{\nu} \propto \nu^{-\alpha}$). The dotted lines correspond to $\alpha=0.6$ and the dashed lines to the steep spectrum with $\alpha=0.85$. Each model shows the characteristic bumps associated to synchrotron emission in the radio and infrared and IC on the CMB to the right of the plot at higher energies. The purple circles represent the fluxes from G279.0+1.1 obtained in this work and the (blue) solid lines are the spectral fit in the LAT range and its $1\sigma$ statistical uncertainty band. The red triangles are the radio fluxes while the cyan line represents a TeV upper limit, both taken from the literature as explained in the text.\label{sed}}
\end{figure}

In the leptonic scenario, a simple model with one population of particles responsible for the synchrotron and IC emission was used. The particle distribution used is a power-law with an exponential cutoff at high energies \citep{1999ApJ...525..368R}. Due to the lack of radio and X-ray observations of the SNR, the broadband spectrum of synchrotron emission is not constrained. \cite{1995MNRAS.277..319D} measured a radio spectral index of $\alpha = 0.6 \pm 0.3$ in a limited frequency range. The corresponding power-law index of the differential gamma-ray spectrum is thus $\Gamma=\alpha +1 = 1.6$, lower than the one measured in this work ($\sim 1.85$) but consistent within uncertainties. Note that this relation is valid below certain energy before reaching any spectral turnover, caused for example by particle cooling. The absence of spectral curvature at GeV energies suggests that the gamma-ray emission is being produced in this regime. In order to explore a range of possible parameters, two versions of the model were applied using radio synchrotron indices of 0.6 and 0.85. The first value was reported by \cite{1995MNRAS.277..319D} as mentioned before, and the second one results from the measured gamma-ray spectral index in this work ($\Gamma = 1.85$). In the absence of data, we also assumed that there are no breaks in the synchrotron spectrum in the radio regime, which have been seen in several SNRs \citep[e.g.,][]{2014Ap&SS.354..541U}. The model and the data are shown in Fig. \ref{sed}.

In the first case ($\Gamma = 1.6$), an average magnetic field of 7 $\mu$G was adopted. The spectrum of the particle distribution has an index of 2.2 and its cutoff energy is 8 TeV. For the corresponding synchrotron flux slope of 0.6 it is hard to reproduce the gamma-ray spectrum and the resulting X-ray fluxes are too high. Although for a distance $d$ to the source in kpc the total energy in electrons above 1 GeV is given by $7.7\times10^{48} \cdot \left( \frac{d}{\mbox{\tiny 3 kpc}} \right)^2$ erg, which is reasonable for a wide range of possible values of $d$. However, due to the lack of X-ray emission indicating that the SNR is old, this scenario for the particle distribution is less likely.

In the other case, for a steep particle distribution with a spectral index of 2.7, the cutoff in the particle energy should be about 25 TeV to be consistent with the LAT fluxes. As can be seen in Fig. \ref{sed}, this value is also consistent with the TeV upper limit. However, we note that the extraction region used for the H.E.S.S. observations is smaller than the size of the SNR itself, and this might indicate that the TeV upper limit is not correct for this source. A cutoff energy of $\sim 25$ TeV could be difficult to achieve in the loss-limited scenario of the standard test-particle theory of diffusive shock acceleration (DSA) for an old SNR. In this scenario, a maximum energy of 25 TeV is obtained with a magnetic field of $\sim \,1$ $\mu$G and a shock speed of 500 km s$^{-1}$, assuming $\eta R_{J} \sim 5$. Here, $\eta$ is a numerical factor related to the magnetic turbulence ($\eta >> 1$ for low turbulence) and $R_{J}$ is related to the shock obliquity \citep[see, e.g.,][]{2008ARA&A..46...89R}. This shock speed is a factor of $\sim 6$ higher than the values derived by \cite{2009MNRAS.394.1791S}. They estimated shock speeds of 75-90 km s$^{-1}$ using the ratios of oxygen lines and H$\beta$ from two filaments. A value of 500 km s$^{-1}$ is high for an old, radiative, SNR, in which case this scenario is not likely. However, these estimates should be treated carefully, as the steep particle distribution required might indicate that test-particle DSA does not apply. In a scenario where the energy is limited by particle escape, the maximum value possible is estimated to be of the order of $\sim 20$ TeV for a similar magnetic field of 1 $\mu$G \citep[e.g.,][]{2008ARA&A..46...89R}. Regardless of the origin of the maximum electron energy, the gamma-ray data can be explained satisfactorily while the resulting X-ray fluxes are low. The total particle energy required is $1.1\times10^{50} \cdot \left( \frac{d}{\mbox{ 3 kpc}} \right)^2$ erg above 1 GeV, for a source distance $d$ in kpc, and the magnetic field is relatively low, 1.6 $\mu$G. The total energy in relativistic electrons in several SNRs has been found to be $\lesssim 1\%$ of the typical kinetic energy available in SNRs, $10^{51}$ erg \citep[see, e.g.,][]{2017ApJ...843...12A,2018A&A...612A...4H}. Therefore, a more reasonable value for G279.0+1.1 would result for an SNR distance lower than 3 kpc. For a distance of 1 kpc to the source, for instance, the total energy in electrons would be of the order of $1\%$ of $10^{51}$ erg, and the corresponding SNR diameter would be $\sim 50$ pc, which is reasonable for an old SNR. With respect to the low magnetic field in the model, as well as the low radio luminosity of this particular SNR, both could be due to expansion in a very low density environment \citep[e.g.,][]{2004A&A...427..525B}. Even if located at the distance estimated to the CO clouds mentioned earlier (3 kpc), there is no indication of interaction of the SNR with these clouds, and the SNR could be expanding in a very low density environment. In any case, the amount of energy in the electrons required by the gamma-ray data favor a distance closer than 3 kpc in the leptonic scenario.

Given this analysis a steep electron index of 2.7 is preferred. The nature of the soft distribution for the particles remains to be discussed. The spectral index of the particles required by the gamma-ray data is very different than that predicted by standard test-particle diffusive shock acceleration. Particle escape could be responsible for the steepening of the gamma-ray emission. Under this hypothesis, the electrons responsible for the radio emission would be located within the actual shell of the SNR (the relatively brighter radio ring), with a synchrotron spectral flux index of $\sim 0.6$, while the softer particle distribution required by the gamma ray observations might result from a population of runaway particles that escaped the SNR in the past. Some models predict the steepening of runaway electrons escaping from SNRs \citep[e.g.,][]{2012MNRAS.427...91O}. This would also explain the fact that the gamma-ray emission region is larger than the ring of radio emission.

An estimate of several timescales becomes relevant in this scenario. The typical energies of electrons upscattering CMB photons to energies of 1--100 GeV are 0.6--6 TeV. The cooling times of electrons with energy $E_e$ due to IC and synchrotron losses are 140 kyr $\cdot \left(\frac{10\,\mbox{\tiny TeV}}{E_e}\right)$ and 320 kyr $\cdot \left(\frac{10\,\mbox{\tiny TeV}}{E_e}\right)$, respectively (assuming a magnetic field value of 1.6 $\mu$G). Therefore IC cooling is slightly faster and, given the energies of the electrons producing the photons detected by the LAT, the cooling times range from 230 kyr to 2.3 Myr. The time when electrons start to escape from an SNR ($t_{esc}$) for typical parameters could range from $10^3$ to $10^4$ yr, depending on the evolution of the magnetic field, according to the model by \cite{2012MNRAS.427...91O}. Assuming that the age of the system is $t \sim 10^5$ yr, then the escape times are much smaller. The runaway electrons could very well survive for a time of $t-t_{esc} \approx \, t$, and thus could propagate a distance of $\sim \sqrt{2 D t}$, where $D$ is the diffusion coefficient. For a typical value of $D$ predicted for the interstellar medium in a standard model for a particle energy of $\sim$ 6 TeV \citep[of the order of $10^{29}$ cm$^2$ s$^{-1}$, see][]{2007ARNPS..57..285S}, this distance is $\sim 260$ pc, which is larger than the radius of the emitting region (73 pc) for a distance to the source of 3 kpc. The diffusion coefficient around an SNR, however, is expected to be lower than the average Galactic value by factors of 10 to 100 due to the amplification of magnetic fluctuations by escaping cosmic rays \cite[e.g.,][and references therein]{2012MNRAS.427...91O,2019MNRAS.490.4317C}. These factors would result in propagation distances of $\sim 80$ to $30$ pc, respectively, which could still be compatible with the extension of the GeV emission seen in the region of G279.0+1.1 given that the true distance to the source is unknown.

It is then possible for electrons to survive long enough and propagate away from the SNR while still producing detectable gamma-ray fluxes. This alternative scenario would be consistent with the steep spectral index observed in the gamma rays, as shown by some simulations. More observations will be needed to determine the real size of the SNR shell at radio wavelengths to confirm this scenario.

Another posibility that would explain the gamma rays from G279.0+1.1 is that the underlying population of particles producing the high-energy gamma rays are protons instead of electrons. Hadronic interactions between accelerated protons and ambient material produce neutral pions that decay to gamma rays with a spectrum that approximately follows the parent particle spectral distribution for homogeneous media. However, the resolution of the gamma-ray data presented in this work is not enough to determine whether the emission comes preferentially from specific regions of the SNR or its surroundings, and a more detailed study should be carried out in the future, including obtaining observations of the ambient gas.

\begin{figure}
\includegraphics[width=9cm,height=5cm]{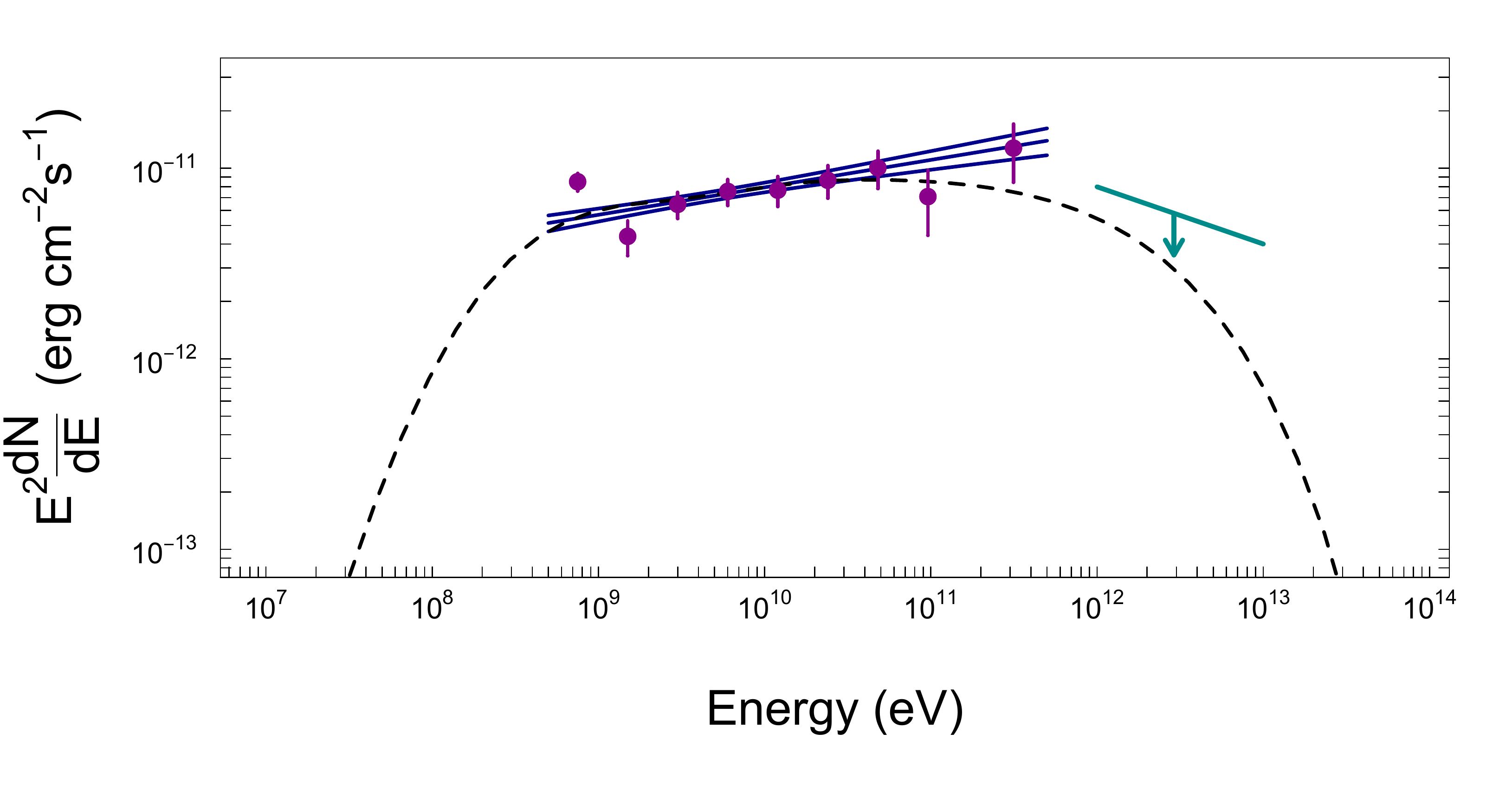}
\caption{The hadronic model for the gamma-ray data from G279.0+1.1 described in the text. The dashed line represents the emission from a cosmic ray population interacting with matter. The data is the same shown in Fig. \ref{sed}.\label{pion}}
\end{figure}

A simple one-zone model was applied to gain an idea of the required parameters. The LAT data cannot constrain the cutoff energy of the particles, and so a fixed value of 25 TeV was adopted as the minimal cutoff energy that predicts fluxes which are still consistent with the highest energy LAT point. The fit to the gamma-ray data points from G279.0+1.1 yields a particle distribution with a power-law spectral index of $2.01\pm0.05$ and a total energy in the protons, in terms of the source distance and the average density of the interacting matter ($n$), of $(3\pm1)\times10^{50} \cdot \left( \frac{d}{\mbox{ 3 kpc}}\right)^2 \cdot \left( \frac{\mbox{ 1 cm}^{-3}}{n} \right)$ erg ($1\sigma$ statistical uncertainties are reported in this case), which is reasonable. This hadronic model is shown in Fig. \ref{pion} on top of the gamma-ray data. The resulting spectral index of the parent particles is consistent with predictions from standard diffusive shock acceleration (DSA) in the test particle limit. Given the relatively hard gamma-ray spectrum, this result is not surprising from the observational point of view. On the other hand, for an evolved SNR, particles outside the SNR could produce a similar spectrum \citep[see, e.g.,][]{2019MNRAS.490.4317C}. More observations are needed to probe the environment in the region, as it has already been pointed out, to conclude anything firmly.

The angular size and previously estimated distance to the SNR (although based on the highly uncertain $\Sigma-D$ relation as well as other uncertain techniques), the high degree of polarization, the optical observations and the presence of old pulsars nearby \citep{1995MNRAS.277..319D}, have led to the classification of G279.0+1.1 as an evolved shell SNR with an age of the order of $10^{5}-10^{6}$ yr. If this is the case, the observation of gamma rays from the source is a bit puzzling. The efficiency of particle acceleration in SNRs is expected to drop as the shock speed decreases. However, recent theoretical work has shown that the spectra of gamma-ray emission from SNRs depend highly on the evironment in which the sources expand, which varies as the remnant evolves and is also influenced by the progenitor of the supernova event \citep{2018MNRAS.475.5237G,Yasuda_2019,2016ApJS..227...28T}. SNRs in low density ($\lesssim 0.1$ cm$^{-3}$) environments with low magnetic fields could show gamma-ray emission that is dominated by IC even if the SNR is old \citep{2012MNRAS.427...91O}. This could be another scenario for the gamma-ray emission seen and more work should be carried out to probe what are the oldest SNRs which could be expected to confine electrons that emit IC gamma rays. There could be a population of remnants evolving in low-density environments with very dim counterparts at other wavelengths, such as HESS J1912+101 \citep{2019RAA....19...45R}. This could also explain the existence of SNR candidates detected only at TeV energies such as HESS J1614-518 \citep{2018A&A...612A...8H}. Other examples of possibly evolved SNRs with hard gamma-ray spectra at GeV energies could be the $\sim 3^{\circ}$-wide SNR G150.8+3.8 which shows dim radio emission \citep{2014A&A...566A..76G,2017ApJ...843..139A}, and a recently discovered source with a very similar size and GeV spectrum, with no known counterpart at lower wavelengths, labeled G350.6-4.7 by \cite{2018MNRAS.474..102A}. However, little is known about these objects and some of them could instead be young SNRs. Deeper observations at different wavelengths are really necessary for these objects and, of course, additional observations of G279.0+1.1 will be important to better constrain its distance and thus age, as well as to probe the environment of the expansion.

\subsection{Summary}
A GeV source has been found coincident with the location of the SNR G279.0+1.1. The gamma-ray emission is $\sim2.8\degr$ wide, which is larger than the previously estimated size of the shell from optical and radio observations \citep{2009MNRAS.394.1791S}. The gamma rays extend farther to the east and north of the radio ring. The spectrum of the gamma rays can be described with a simple power-law of the form $\frac{dN}{dE} \propto E^{-\Gamma}$ with $\Gamma \sim 1.85$.

The leptonic scenario for the gamma rays from G279.0+1.1 favors a distance smaller than 3 kpc to the source, a particle distribution that is considerably steeper than predictions from standard, test-particle, shock acceleration theory, and a low magnetic field ($\sim 1.6$ $\mu$G). This results in low synchrotron fluxes at all wavelengths from the particles responsible for the gamma rays. A simple one-zone model of electrons was used and no spectral breaks are assumed in the calculation of the fluxes.

In another plausible leptonic scenario, which would account for the possible different slopes seen in the radio and gamma ray bands (and particularly the steep gamma-ray spectrum), the radio ring seen indeed represents the entirety of the shell of the SNR, where the synchrotron emission is produced, while the gamma rays are caused by escaping electrons. This also explains the different sizes of the two regions. Deeper observations in the radio and gamma-ray bands are highly encouraged for this SNR.

The hadronic scenario requires a particle spectral index of 2. This scenario for the gamma rays is reasonable from the point of view of the energy needed. Even for a distance of 3 kpc and a total particle energy content of $\sim 10^{50}$ erg, the required average density in the target material is $\sim3$ cm$^{-3}$, which is not uncommon. More observations with gamma-ray instruments with better resolution as well as molecular and gas studies are encouraged for the SNR G279.0+1.1.

\section*{Acknowledgements}

The author would like to thank Natasha Hurley-Walker for pointing out the existence of a radio survey used in this work. The comments made by the anonymous referee helped improve the quality of this work considerably. This project has received funding from the European Union's Horizon 2020 research and innovation programme under the Marie Sk\l{}odowska-Curie grant agreement No 690575 and from Universidad de Costa Rica and its Escuela de F\'isica.

\bibliographystyle{mnras}
\bibliography{references}

% Don't change these lines
\bsp    % typesetting comment
\label{lastpage}

\end{document}